\pdfoutput=1
\documentclass{llncs}

\usepackage{todonotes}
\usepackage{amsmath,amsfonts,amssymb}
\usepackage{stmaryrd}
\usepackage{xspace}
\usepackage{booktabs}
\usepackage{tabularx}
\usepackage{todonotes}
\usepackage{wrapfig}
\usepackage[ruled,linesnumbered,algo2e]{algorithm2e}
\usepackage{float}
\restylefloat{table}
\usepackage{url}
\usepackage{graphicx}
\usepackage{tikz}
\usetikzlibrary{fit,shapes}
\usepackage{multirow}
\usepackage{hhline}
\usepackage{array}
\usepackage{enumerate,paralist}
\usepackage{eurosym}
\usepackage{hyperref}
\usepackage{colortbl}
\definecolor{incolor}{RGB}{210,220,230}
\definecolor{outcolor}{RGB}{235,215,215}

\title{Semantics and Analysis of DMN Decision Tables}

\author{Diego Calvanese$^1$ \and Marlon Dumas$^2$ \and \"Ulari Laurson$^2$,
\\  Fabrizio M.~Maggi$^2$ \and Marco Montali$^1$ \and Irene Teinemaa$^2$}
\authorrunning{D.~Calvanese, M.~Dumas, \"U.~Laurson, F.M.~Maggi,
  M.~Montali, I.~Teinemaa}

\institute{
$^1$Free University of Bozen-Bolzano, Italy\\
$^2$University of Tartu, Estonia\\
}

\usetikzlibrary{positioning}

\newcommand{\ours}{our approach}
\newcommand{\theirs}{Signavio}

\newcommand{\true}{\mathsf{true}}
\newcommand{\false}{\mathsf{false}}

\newcommand{\dl}{\ensuremath{\D}\xspace}

 \newcommand{\D}{\mathcal{D}}

\newcommand{\defterm}[1]{\mbox{\underline{\it\smash{#1}\vphantom{\lower.1ex\hbox
{x}}}}}

\newcommand{\set}[1]{\{#1\}}                      % set

\newcommand{\tup}[1]{\langle #1\rangle}            % tuple

\newcommand{\Mod}[1]{\mathit{Mod}}

\newcommand{\qedboxfull}{\vrule height 5pt width 5pt depth 0pt}
\newcommand{\qedfull}{\hfill{\qedboxfull}}

\newcolumntype{C}{>{\centering\arraybackslash}X}
\newcolumntype{R}{>{\raggedleft\arraybackslash}X}
\newcolumntype{L}{>{\raggedright\arraybackslash}X}

\begin{document}

\maketitle

\begin{abstract}
The Decision Model and Notation (DMN) is a standard notation to capture decision logic in business applications in general and business processes in particular. A central construct in DMN is that of a decision table. The increasing use of DMN decision tables to capture critical business knowledge raises the need to support analysis tasks on these tables such as correctness and completeness checking. This paper provides a formal semantics for DMN tables, a formal definition of key analysis tasks and scalable algorithms to tackle two such tasks, i.e., detection of overlapping rules and of missing rules. The algorithms are based on a geometric interpretation of decision tables that can be used to support other analysis tasks by tapping into geometric algorithms. The algorithms have been implemented in an open-source DMN editor and tested on large decision tables derived from a credit lending dataset.
\end{abstract}


\section{Introduction}
\label{sec:intro}

Business process models often encode decision logic of varying complexity, typically via conditional expressions attached either to outgoing flows of decision gateways or to conditional events. The need to separate this decision logic from the control-flow logic~\cite{Batoulis2015} and to capture it at a higher level of abstraction has motivated the emergence of the Decision Model and Notation (DMN)~\cite{DMN}.

A central construct of DMN is that of a decision table, which stems from the notion of decision table proposed in the context of program decision logic specification in the 1960s~\cite{Pooch74}. A DMN decision table consists of columns representing the inputs and outputs of a decision, and rows denoting rules. Columns may be typed, meaning that they have an associated domain (or facet). Each rule is a conjunction of basic expressions captured in an expression language known as S-FEEL (Simplified Friendly Enough Expression Language).

The use of DMN decision tables as a specification vehicle for critical business decisions raises the question of ensuring the correctness of these tables, in particular the detection of inconsistent or incomplete DMN decision tables. Indeed, detecting errors in DMN tables at specification time may prevent costly defects down the road during business process implementation and execution.

This paper provides a foundation for analyzing the correctness of DMN tables. The contributions of the paper are: (i) a formal semantics of DMN tables; (ii) a formalization of correctness criteria for DMN tables; and (iii) scalable algorithms for two basic correctness checking tasks over DMN tables, i.e., detection of overlapping rules and detection of missing rules (i.e., incompleteness). The latter algorithms are based on a novel geometric interpretation of DMN tables, wherein each rule in a table is mapped to an iso-oriented hyper-rectangle in an N-dimensional space (where N is the number of columns). Accordingly, the problem of detecting overlapping rules is mapped to that of detecting overlapping hyper-rectangles. Meanwhile, the problem of detecting missing rules is mapped to that of differencing the N-dimensional universe defined by the N columns of a DMN table, and the set of hyper-rectangles induced by its rules. Based on this geometric interpretation and inspired by sweep-based spatial join algorithms~\cite{Arge1998}, the paper presents scalable algorithms for these two analysis tasks. The algorithms have been implemented atop the \textsf{dmn-js} DMN editor and evaluated over decision tables of varying sizes derived from a credit lending dataset.

The rest of the paper is structured as follows. Section~\ref{sec:back} introduces DMN and discusses related work. Section~\ref{sec:formal} presents the formalization of DMN tables and their associated correctness criteria. Section~\ref{sec:algo} presents the algorithms for correctness analysis while Section~\ref{sec:exper} discusses their empirical evaluation. Finally, Section~\ref{sec:conclusion} summarizes the contributions and outlines future work directions.

\section{Background and Related Work}
\label{sec:back}
%Below, we provide an overview of DMN decision tables and discuss related work.

\subsection{Overview of DMN Decision Tables}
A DMN table consists of columns corresponding to input or output attributes, and rows corresponding to rules. Each column is associated to a type (e.g., a string, a number, or a date), and optionally to a more specific domain of possible values, which we hereby call a \emph{facet}. Each row has an identifier, one expression for each input column (a.k.a.\ the \emph{input entries}), and one specific value for each output column (the \emph{output entries}). For example, Table 1 shows a DMN table with two input columns, one output column and four rules.
%\footnote{DMN supports different equivalent graphical renderings for the same decision, but this aspect is orthogonal to our contribution.}

%rule has a number and a set of

%for a reference to the graphical DMNnotation)

% \begin{figure}[hbt]
% \centering
% \includegraphics[width=.75\textwidth]{imgs/Decision_Table}
% \label{fig:decision-table}
% \caption{Sample decision table with its constitutive elements}
% \end{figure}

\newcounter{nodecount}
% Command for making a new node and naming it according to the nodecount counter
\newcommand\tabnode[1]{\addtocounter{nodecount}{1} \tikz \node (\arabic{nodecount}) {#1};}

\newcommand{\verygood}{\texttt{VG}}
\newcommand{\good}{\texttt{G}}
\newcommand{\fair}{\texttt{F}}
\newcommand{\poor}{\texttt{P}}

 \tikzstyle{every picture}+=[remember picture,baseline]
 \tikzstyle{every node}+=[inner sep=0pt,anchor=base,
 minimum width=.3cm, minimum height=.3cm,align=center,text depth=0ex,outer sep=0pt]
\tikzstyle{legend}=[rectangle,thin,draw=black!75,fill=orange!20,minimum height=5mm,minimum width=2.1cm,inner sep=2pt]

\begin{table}[hbt]
\centering
\begin{tabularx}{.5\textwidth}{|c|C|C||C|}
\cline{1-2}
\multicolumn{2}{|c|}{\tabnode{\textbf{Loan Grade}}}\\
\hline
\tabnode{\textbf{U}}\tabnode{\textbf{C}}~
&
\cellcolor{incolor}
\tabnode{\textbf{Annual}}
&
\cellcolor{incolor}
\tabnode{\textbf{Loan}}
&
\cellcolor{outcolor}
\tabnode{\textbf{Grade}}\\
&
\cellcolor{incolor}
\textbf{Income}
&
\cellcolor{incolor}
\textbf{Size}
&
\cellcolor{outcolor}
\\
\cline{2-4}
&
$\geq 0$
&
$\geq 0$
&
\tabnode{\verygood,\good,\fair,\poor}\\
\hline
A
&
$[0..1000]$
&
$[0..1000]$
&
\verygood\\
\hline
 B
&
$[250..750]$
&
$[4000..5000]$
&
\good\\
\hline
 C
&
$[500..1500]$
&
$[500..3000]$
&
\fair\\
\hline\tabnode{D}~
&
\tabnode{$[2000..2500]$}
&
\tabnode{$[0..2000]$}
&
\tabnode{\poor}\\
\hline
\multicolumn{1}{c}{}\\
\end{tabularx}
\begin{tikzpicture}[overlay,node distance=1mm and 8mm]
\node[legend, left = of 1] (name) {Table name};
\node[legend, below = of name] (hi) {Hit indicator};
\node[legend, below = of hi,yshift=-.2cm] (ci) {Completeness\\ indicator};
\node[legend, above = of 5,xshift=1cm] (in) {Input attrs};
\node[legend, right = of 7] (fac) {Facet};
\node[legend, above = of fac] (out) {Output attr};
\node[legend, below = of fac,yshift=-.2cm] (rule) {Rule};

\node[legend, below = of ci,yshift=-.37cm] (pi) {Priority\\ indicator};
\node[legend, below = of 10,xshift=-1cm,yshift=-.1cm] (ie) {Input entries};
\node[legend, below = of 11,yshift=-.1cm] (oe) {Output entry};

\draw[->] (name) edge (1);
\draw[->] (hi) edge (2);
\draw[->] (ci) edge (3);
\draw[->] (in) edge (4);
\draw[->] (in) edge (5);
\draw[->] (out) edge (6);
\draw[->] (fac) edge (7);
\draw[->] (rule) edge (-0.11,-0.29);
\draw[->] (pi) edge (8);
\draw[->] (ie) edge (9);
\draw[->] (ie) edge (10);
\draw[->] (oe) edge (11);

\end{tikzpicture}
\label{fig:decision-tab}
\caption{Sample decision table with its constitutive elements}
\end{table}

\vspace{-0.3cm}
Given an input configuration consisting of a vector of values (one entry per column), if every input entry of a row holds true for this input vector, then the vector \emph{matches} the row and the output entries of the row are evaluated. For example, vector $\langle500, 4230\rangle$ matches rule $B$ in Table 1, thus yielding $\good$ in the output configuration.
To specify how output configurations are computed from input ones, a DMN table may have a \emph{hit indicator} and a \emph{completeness indicator}. The hit indicator specifies whether only one or multiple rows of the table may match a given input, and if multiple rules match an input, how should the output configuration be computed. The completeness indicator specifies whether every input configuration must match at least one rule or potentially none. If an input configuration matches two or more rules, this may contradict the hit policy. Similarly, if no rule matches an input configuration, this may contradict the completeness indicator. The former type of contradiction is called \emph{overlapping rules} while the latter is called \emph{missing rule}.

\vspace{-0.2cm}
\subsection{Analysis of DMN Decision Tables}

%This approach has the advantage that it generalizes well to the case of uncertain information.
The need to analyze decision tables from the perspective of completeness (i.e., detecting missing rules) as well as consistency and non-redundancy (i.e., detecting overlapping rules) is widely recognized~\cite{CodasylReport}. These two analysis tasks have been tackled using rough sets~\cite{Pawlak1987}. However, this approach requires that the domains of the input attributes are boolean or categorical. Numerical attributes need to be previously discretized into intervals. In this paper, we study the problem of analyzing decision tables with arbitrary S-FEEL expressions, meaning that no prior discretization of numerical domains is required.

Prologa~\cite{Vanthienen1994,Vanthienen1998} is a tool for modeling and executing classical decision tables. It supports the construction of decision tables in a way that prevents overlapping or missing rules. It also supports the optimization of a decision table via rule merging: two rules are merged when all but one of their input entries are identical, and their output entries are also identical. However, Prologa presents the same intrinsic limitation of the rough set approach: it requires columns to have boolean or categorical domains. Hence, numerical domains need again to be discretized into intervals when constructing a decision table.
% (i.e., in which case the one input where the two rules differ can be merged into a single expression and thus the columns can be merged)

Signavio's DMN editor\footnote{\url{http://www.signavio.com}} detects overlapping and missing rules. However, the employed analysis techniques are undisclosed and no empirical evaluation thereof has been reported. Also, the diagnosis of overlapping and missing rules produced by Signavio is unnecessarily large: it often reports the same rule overlap multiple times. This behavior will be further explained in Section~\ref{sec:exper}.
%.detects and reports overlaps along each column rather than considering all columns at once

OpenRules\footnote{\url{http://openrules.com/}} uses constraint satisfaction techniques to analyze business rules, in particular rules encoded in decision tables. While using a general solver to analyze decision tables is an option (e.g., an SMT solver such as Z3~\cite{Z3}), this approach leads to a boolean output (is the set of rules satisfiable?), and cannot natively highlight specific sets of rules that need to be added to a table (missing rules), nor specific overlaps between pairs of rules that need to be resolved.

\newcommand{\dt}{\mathcal{D}}
\newcommand{\cval}[1]{\mathtt{#1}}
\newcommand{\func}[1]{\mathsf{#1}}
\newcommand{\attr}[1]{\mathbf{#1}}
\newcommand{\formula}[1]{\mathit{#1}}
\newcommand{\tname}{T}
\newcommand{\tcompl}{C}
\newcommand{\thit}{H}
\newcommand{\tpriority}{\func{Priority}}
\newcommand{\complete}{\cval{c}}
\newcommand{\incomplete}{\cval{i}}
\newcommand{\upol}{\cval{u}}
\newcommand{\apol}{\cval{a}}
\newcommand{\ppol}{\cval{p}}
\newcommand{\fpol}{\cval{f}}
\newcommand{\opol}{\cval{o}}
\newcommand{\rpol}{\cval{r}}
\newcommand{\cpol}{\cval{c}}
\newcommand{\tin}{I}
\newcommand{\tout}{O}
\newcommand{\atype}{\func{Type}}
\newcommand{\afacet}{\func{Facet}}
\newcommand{\types}{\mathfrak{T}}
\newcommand{\type}{\mathcal{T}}
\newcommand{\dom}{\Delta}
\newcommand{\sig}{\Sigma}
\newcommand{\sigp}{\sig^P}
\newcommand{\sigf}{\sig^F}
\newcommand{\cond}{\mathcal{Q}}
\newcommand{\anycond}{\mathtt{-}}
\newcommand{\orcond}{\mathtt{,}}
\newcommand{\trules}{{R}}
\newcommand{\incond}{\func{If}}
\newcommand{\outval}{\func{Then}}
\newcommand{\tpr}{\func{Priority}}
\newcommand{\cf}[1]{\Phi_{#1}}

\newcommand{\legal}[2]{\formula{Legal}_{#1}(#2)}
\newcommand{\comp}[2]{\formula{Compatible}_{#1}^{#2}}
\newcommand{\matches}[3]{\formula{Matches}_{#1}^{#2}(#3)}
\newcommand{\adm}[1]{\formula{Admissible}_{#1}}
\newcommand{\correct}[1]{\formula{Correct}_{#1}}
\newcommand{\trig}[2]{\formula{TriggeredBy}_{#1}(#2)}
\newcommand{\iorel}[3]{\formula{IORel}_{#1}(#2,#3)}
\newcommand{\compltable}[1]{\formula{Complete}_{#1}}
\newcommand{\uniquetable}[1]{\formula{Unique}_{#1}}
\newcommand{\anytable}[1]{\formula{AgreesOnOutput}_{#1}}
\newcommand{\ptrig}[2]{\formula{TriggeredWithPriorityBy}_{#1}(#2)}
\newcommand{\uaiorel}[3]{\formula{IORelUA}_{#1}(#2,#3)}
\newcommand{\piorel}[3]{\formula{IORelP}_{#1}(#2,#3)}
\newcommand{\maskedby}[2]{\formula{MaskedBy}_{#1}^{#2}}
\newcommand{\hasirr}[1]{\formula{HasIrrelevantRules}_{#1}}

\section{Formalization}
\label{sec:formal}
In this section, we provide a logic-based formalization of DMN
decision tables, unanmbiguously defining their input/output semantics,
and at the same time introducing several analysis tasks focused on
correctness checking. As a concrete
specification language for input entries in DMN, we consider
the S-FEEL language, introduced in the DMN standard
itself. 

Our formalization is based on classical predicate logic extended with
data types, which are needed to capture
conditions that employ domain-specific predicates such as comparisons
interpreted over the total order of natural numbers. Such
formalization is important per s\`e, as it defines a clear,
unambiguous semantics of decision tables, and also as an interlingua
supporting the comparison of different analysis techniques.

\subsection{Data Types and S-FEEL Conditions}
We first introduce the building blocks of decision tables,
i.e., the types of the modeled attributes, and conditions over such
types expressed using the S-FEEL language.
% \begin{table}[t]
% \begin{center}
% \begin{tabularx}{.48\textwidth}{|c|c|c|c||c|c|c|}
% \cline{1-2}
% \multicolumn{2}{|l|}{\textbf{Routing rules}}\\
% \hline
% \textbf{P}
% &
% \cellcolor{incolor} Applicant Age
% &
% \\
% \end{tabularx}
% \end{center}
% \end{table}
A data type $\type$ is a tuple
$\tup{\dom_\type,\sig_\type}$, where $\dom_\type$ is an \emph{object domain},
and $\sig_\type = \sigp_\type \uplus \sigf_\type$ is a \emph{signature},
constituted by a set $\sigp_\type$ of \emph{predicate
  symbols}, and a set $\sigf_\type$ of \emph{function symbols}
(disjoint from $\sigp_\type$).
Each predicate symbol $R \in \sigp_\type$ comes with its own arity $n$, and with an
$n$-ary predicate $R^\type \subseteq \dom_\type^n$ that rigidly defines its
semantics. Each function symbol $f \in \sigf_\type$ comes with its own
arity $m$, and with a function $\dom_\type^m \rightarrow \dom_\type$
that defines its semantics. To make the arity explicit in predicate
and function symbols, we use the standard notation $R/n$ and $f/m$.
As usual, we assume that every data type is equipped
\emph{equality} as a predefined, binary predicate interpreted as the
identity on the underlying domain. Hence, we will not explicitly
mention equality in the signatures of data types.
%
% \begin{example}
% Consider the data type
% $\tup{\mathbb{Z},\set{\mathbf{0},\mathbf{1},+,-,\leq}}$, where
% $\mathbf{0}$ and $\mathbf{1}$ are nullary, while $+$, $-$ and $\leq$
% are binary. When such predicate symbols are interpreted in the usual
% way over the integers, this data type corresponds to the
% \emph{Presburger arithmetic}. \qedfull
% \end{example}
In the following, we show some of the S-FEEL data
types\footnote{Date/time data types are also supported, but from the
  logical point of view they can be considered as simple numeric attributes}:
\begin{compactitem}
\item $\type_{\mathbb{S}} = \tup{\mathbb{S},\emptyset,\emptyset}$ --
  strings.
\item $\type_{\mathbb{B}} =
  \tup{\set{\true,\false},\emptyset,\emptyset}$ -- boolean
  attributes.
\item $\type_{\mathbb{Z}} = \tup{\mathbb{Z},
    \set{\mathbf{0}/0,\mathbf{1}/0,{<}/2,{>}/2},\set{{+}/2,{-}/2,{\cdot}/2,{\div}/2}}$
  -- integer numbers equipped with the usual comparison predicates and
  binary operations;
\item $\type_{\mathbb{R}}$ (defined as $\type_{\mathbb{Z}}$ by
  replacing the domain $\mathbb{Z}$ with $\mathbb{R}$, and by
  reinterpreting all predicates and functions accordingly) -- real numbers
  equipped with the usual comparison predicates and binary operations.
\end{compactitem}
The set of all such types is
denoted by $\types$. Since decision tables do not support
conditions that combine multiple data types, we can safely assume that
the \emph{object domains of all types in $\types$ are pairwise
disjoint}.%  It is important to notice that all types in $\types$ are
% equipped with binary predicates only, among which there is (domain-specific) equality.

S-FEEL allows one to formulate conditions over types. Such conditions
constitute the basic building blocks for facets and rules, which in
turn are the core of decision tables. The syntax of an
\emph{(S-FEEL) condition} $\cond$ over type is:
\\
$
\begin{array}{rcl}
\cond
&::= &
``\anycond"
\mid
\mathit{Term}
\mid
\mathtt{``not("}
~Term~
\mathtt{``)"}
\mid
\mathit{Comparison}
\mid
\mathit{Interval}
\mid
\cond_1 \orcond \cond_2 \\
\mathit{Comparison}
&::=&
\mathit{COp}~\mathit{Term}
\\
\mathit{COp}
&::=&
``{<}" \mid ``{>}" \mid ``{\leq }" \mid ``{\geq}"
\\
\mathit{Interval}
&::=&
(``(" \mid ``[")
~
Term_1
~
``\mathtt{..}"
~
Term_2
~
(``)" \mid ``]")
\\
\mathit{Term}
&::= &
v
\mid
f(\mathit{Term}_1,\ldots,\mathit{Term}_m)
\end{array}
$
where
%\begin{inparaenum}[\it (i)]
%\item $P$ is a binary predicate symbol in $\sig_\type$,
%\item
$v$ is an object %from $\dom_\type$,
%\item
and $f$ is an $m$-ary function. %in $\sig_\type$.
%\end{inparaenum}

Intuitively,
S-FEEL supports the following conditions on a given data type $\type = \tup{\dom_\type,\sig_\type}$:
\begin{inparaenum}[\it (i)]
\item ``$\anycond$'' indicates \emph{any value}, i.e., it holds for every
  object in $\dom_\type$.
\item $\mathit{Term}$ is a shortcut for ``$= \mathit{Term}$'', and
  indicates a \emph{matching expression}, which holds for the
  object in $\dom_\type$ that corresponds to the result denoted by term
  $\mathit{Term}$. A term, in turn, corresponds either to a specific object in
  $\dom_\type$, or to the recursive application of an $m$-ary function in
  $\sig_\type$ to $m$ terms.
\item $\mathit{Comparison}$ is only applicable when $\type$ is a
  numeric data type, and indicates a \emph{comparison condition},
 which holds for all objects that are related via the employed
 comparison predicate to the object
  resulting from expression $\mathit{Term}$.
\item $\mathit{Interval}$ is only applicable when $\type$ is numeric,
  and allows the modeler to capture membership conditions that tests
  whether an input object belongs to the modeled interval.
\item ``$\cond_1 \orcond \cond_2$'' indicates an \emph{alternative
    condition}, which holds whenever one of the two conditions
  $\cond_1$ and $\cond_2$ holds.
\end{inparaenum}

\begin{example}
\label{ex:facets}
The fact that a risk category is either high, medium or
low can be expressed by the following condition over
$\type_\mathbb{S}$:
``$\cval{high}\orcond\cval{medium}\orcond\cval{low}$''.
By using $\type_{\mathbb{Z}}$ to denote the age of persons (in years),  the group of
people that are underage or old (i.e., having at least $70$ years) is
captured by condition ``$[0..18]\orcond\geq 70$''.
\qedfull
\end{example}

\subsection{Decision Tables}
We are now in the position of defining DMN decision tables. See
Table 1 for a reference example.
A \emph{decision table} $\dt$ is a tuple
$\tup{\tname,\tin,\tout,\atype,\afacet,\trules,\tpr,\tcompl,\thit}$, where:
\begin{compactitem}
\item $\tname$ is the \emph{table name}.
\item $\tin$ and $\tout$ are disjoint, finite sets of \emph{input} and
  \emph{output attributes} (represented as strings).\footnote{These
    are called ``expressions'' in the DMN standard, but we prefer the term
  ``attribute'' as it is less ambiguous.}
\item $\atype: \tin \uplus \tout \rightarrow \types$ is a \emph{typing
    function} that associates each input/output attribute to its corresponding data
  type. %The type is not explicitly shown graphically in the DMN notation.
\item $\afacet$ is a \emph{facet function} that associates each
  input/output attribute $\attr{a} \in \tin \uplus \tout$ to a condition over
  $\atype(\attr{a})$, defining the \emph{acceptable objects} for
  that attribute. Facet functions are depicted as ``optional lists of
  values'' in Table 1.
\item $\trules$ is a finite set of \emph{rules}
  $\set{r_1,\ldots,r_p}$. Each rule $r_k$ is a pair
  $\tup{\incond_k,\outval_k}$, where $\incond_k$ is an \emph{input entry function} that
  associates each input attribute $\attr{a^{in}} \in \tin$ to a condition over
  $\atype(\attr{a^{in}})$, and $\outval_k$ is an \emph{output entry function} that
  associates each output attribute $\attr{a^{out}} \in \tout$ an object in
  $\atype(\attr{a^{out}})$.
\item $\tpr : \trules \rightarrow \set{1,\ldots,|\trules|}$ is a \emph{priority
    function} injectively mapping rules in $\tpr$ to a corresponding
  rule number defining its priority. If
  no priority is explicitly given, in accordance with the standard we assume that the priority is
  implicitly defined by the graphical ordering in which rule entries
  appear inside the decision table.
\item $\tcompl \in \set{\complete,\incomplete}$ is the
  \emph{completeness indicator}, where $\complete$ is the
  default value and stands
  for \emph{complete} table, while $\incomplete$ stands for
  \emph{incomplete} table.
\item $\thit \in \set{\upol,\apol,\ppol,\fpol}$ % ,\opol,\rpol,\cpol}$
is the \emph{(single) hit indicator} defining the policy for the rule
application, where:
\begin{inparaenum}[\it (i)]
\item $\upol$ is the default value and stands for \emph{unique hit policy},
\item $\thit = \apol$ stands for \emph{any hit policy},
\item $\thit = \ppol$ stands for \emph{priority hit policy}, and
\item $\thit = \fpol$ stands for \emph{first hit policy}.
\end{inparaenum}
\end{compactitem}
\smallskip
We now informally review the intuitive semantics of rules and of
completeness/hit indicators in DMN, moving to the formalization in Section~\ref{sec:reasoning}.

\smallskip
\noindent
\textbf{Rule semantics.}
Intuitively, rules follow the standard ``if-then''
interpretation. Rules are matched against \emph{input configurations}, which map the
input attributes to objects in such a way that each object
\begin{inparaenum}[\it (i)]
\item belongs to the type of the corresponding input attribute, and
\item satisfies the corresponding facet.
\end{inparaenum}
If, for every input attribute, the
assigned object satisfies the condition imposed by the rule on that
type, then the rule \emph{triggers}, and bounds the output attributes
to the actual objects mentioned by the rule.

\begin{example}
Consider the decision table in Table 1. The
input configuration where $\attr{Income}$ is $\cval{500}$ and
$\attr{Loan}$ is $4230$, triggers rule $B$. \qedfull
\end{example}

\smallskip
\noindent
\textbf{Completeness indicator.}
When the table is declared to be complete, the intention is that every
possible input configuration must trigger at least one rule. Incomplete
tables, instead, have input configurations with
no matching rule.

\smallskip
\noindent
\textbf{Hit policies.}
%how to match input configurations to a single configuration of output objects, also indicating
Hit policies specify how to handle the case where multiple
rules are triggered by an input configuration. In particular:
\begin{compactitem}
\item ``Unique hit'' indicates that at most one rule can be triggered by a
  given input configuration, thus avoiding the need of handling how to
  compute the output objects in the case of multiple triggered rules.
\item ``Any hit'' indicates that when multiple rules are triggered, they
  must agree on the output objects, thus guaranteeing that the output is
  unbambiguous.
\item ``Priority hit'' indicates that whenever multiple rules trigger,
  then the output is unambiguously computed by only considering the
  contribution of the triggered rule that has highest priority.
\item ``First hit'' can be understood as a variant of the priority hit, in
  which priority is implicitly obtained from the ordering in which
  rules appear in the decision table. Hence, this
  case is subsumed by that of priority hit.
  \item ``Collect'' implies that multiple rules can match an input configuration and when this is the case, all matching rules are fired the the resulting output configurations are aggregated. Aggregation is orthogonal to correctness checking, and thus we leave the ``Collect'' policy outside the scope of the formalization below.
\end{compactitem}

%Note that a decision table may violate its hit policy (e.g.\ a decision table with unique hit policy may contain overlapping rules) or it may violate its completeness policy (e.g.\ a ``complete'' table may have missing rules).
%This shows that DMN must be equipped with dedicated analysis tasks tailored to the detection of such potential discrepancies.

\subsection{Formalization of Rule Semantics and of Analysis Tasks}
\label{sec:reasoning}

We first define how conditions map to corresponding
formulae. Since each condition is applied to a single input
attribute, the corresponding formula has a single free variable
corresponding to that attribute.
Given a condition $\cond$ over
type $\type = \tup{\dom_\type,\sig_\type}$, the \emph{condition formula for
  $\cond$}, written $\cf{\cond}$, is a formula using
predicates/functions in $\sig_\type$ and objects from $\dom_\type$,
and possibly mentioning a single free variable, constructed as follows:
\\
$
\cf{\cond} \triangleq
\begin{cases}
\mathit{true} & \text{if } \cond = ``\anycond"\\
\neg \Phi_{\mathit{Term}}
& \text{if } \cond =
``\mathtt{not(}\mathit{Term}\mathtt{)}"\\
x = \mathit{Term}  & \text{if } \cond = \mathit{Term}\\
x~COp~\mathit{Term} & \text{if } \cond = ``
\mathit{COp}~\mathit{Term}"  \text{and } \mathit{COp} \in \set{<,>,\leq,\geq}\\
x > \Phi_{\mathit{Term_1}} \land x < \Phi_{\mathit{Term_2}}
&
\text{if } \cond = ``(Term_1..Term_2)"
\\
x > \Phi_{\mathit{Term_1}} \land x \leq \Phi_{\mathit{Term_2}}
&
\text{if } \cond = ``(Term_1..Term_2]"
\\
x \geq \Phi_{\mathit{Term_1}} \land x < \Phi_{\mathit{Term_2}}
&
\text{if } \cond = ``[Term_1..Term_2)"
\\
x \geq \Phi_{\mathit{Term_1}} \land x \leq \Phi_{\mathit{Term_2}}
&
\text{if } \cond = ``[Term_1..Term_2]"
\\
\cf{\cond_1}{x} \lor \cf{\cond_2}{x} & \text{if } \cond = ``\cond_1 \orcond \cond_2"\\
\end{cases}
$
As usual, we also use notation $\cf{\cond}(x)$ to explicitly mention
the free variable of the condition formula.

\begin{example}
Consider the S-FEEL conditions in Example~\ref{ex:facets}. The
condition over the risk category is
$\mathit{Risk} = \cval{high} \lor \mathit{Risk} =
\cval{medium} \lor \mathit{Risk} =
\cval{low}$.
The condition formula person ages is instead: $(\mathit{Age} \geq 0
\land \mathit{Age} \leq 18) \lor \mathit{Age} \geq 70$. \qedfull
\end{example}

% \todo[inline]{For the moment, we abstract away from SMT and the
%   theory-dependent notions of entailments. We just focus on the
%   construction of the formulae capturing the semantics of rules and of
% reasoning tasks. Notice that we are in a very simple case, as there
% are no interface variables. Notice also that, due to this, we can
% consider all signatures and data types to be pairwise disjoint. This
% is also something that helps a lot. We have to understand more about
% the implications of these properties.}

With this notion at hand, we now formalize the notions of correctness of rule specifications, semantics of rules,
and semantics of completeness and hit indicators. These notions are building blocks for an overall notion of \emph{table
  correctness}.

Let $\dt =
\tup{\tname,\tin,\tout,\atype,\afacet,\trules,\tpr,\tcompl,\thit}$ be
a decision table with $m$ input attributes $\tin =
\set{\attr{a_1},\ldots,\attr{a_m}}$, $n$ output attributes $\tout =
\set{\attr{b_1},\ldots,\attr{b_n}}$, and $p$ rules $\trules = \set{r_1,\ldots,r_p}$.
% , where:
% \begin{inparaenum}[\it (i)]
% \item;
% \item $\tout = \set{\attr{a_1^{out}},\ldots,\attr{a_n^{out}}}$;
% \item $\trules = \set{r_1,\ldots,r_p}$.
% \end{inparaenum}
We use variables $x_1,\ldots,x_m$ for objects matching the input attributes,
and variables $y_1,\ldots,y_n$ for those matching the output attributes.

\smallskip
\noindent
\textbf{Facet correctness.}
We first consider the \emph{Facet correctness} of  $\dt$, which
intuitively amounts to check whether all the mentioned
input conditions and output objects are compatible with their
corresponding attribute facets.

Given an attribute $\attr{a} \in
\tin \cup \tout$ and a corresponding input variable $x$, the fact that
\emph{$x$ is legal for $\attr{a}$} is defined as:
\[
\legal{\attr{a}}{x} \triangleq \cf{\afacet(\attr{a})}(x)
\]
We use this notion in combination with a condition $\cond$ over
$\attr{a}$, so as to check whether an input variable \emph{$x$ matches
with $\cond$}:
\[
\matches{\attr{a}}{\cond}{x} \triangleq \legal{\attr{a}}{x} \land \cf{\cond}(x)
\]
Note that for output objects, $\cf{\cond}(x)$ above is a test
where $x$ is equated to the output object.

This derived predicate, in turn, can be used to identify whether
\emph{$\cond$ is compatible with $\attr{a}$}, i.e., whether the
condition is specified in such a way that can potentially trigger, or
is instead contradictory with the facet attached to $\attr{a}$:
\[
\comp{\attr{a}}{\cond} \triangleq \exists x. \matches{\attr{a}}{\cond}{x}
\]

\smallskip
\noindent
\textbf{Rule semantics.}
A rule $r = \tup{\incond,\outval} \in \trules$
  is \emph{triggered by} a configuration $x_1,\ldots,x_m$ of input
  objects whenever each such object matches with the corresponding
  input condition:
\[
\trig{r}{x_1,\ldots,x_m} \triangleq
\bigwedge_{i \in \set{1,\ldots,m}} \matches{\attr{a_i}}{\incond(\attr{a_i})}{x_i}
\]
Two configurations $\vec{x}$ and $y_1,\ldots,y_n$ of input and
output objects are \emph{input-output related} by a rule $r =
\tup{\incond,\outval} \in \trules$ if the rule is triggered by the
input configuration, and binds the output as specified by the output
configuration:
\[
\iorel{r}{\vec{x}}{y_1,\ldots,y_n} \triangleq
\trig{r}{\vec{x}} \land \bigwedge_{j \in \set{1,\ldots,n}} \matches{\attr{b_j}}{\outval(\attr{b_j})}{y_j}
\]

\smallskip
\noindent
\textbf{Completeness.}
When declaring that a table is (in)complete, there is no guarantee
that the specified rules guarantee this property. To check whether
this is indeed the case, we introduce a formula that holds whenever
each possible input configuration triggers at least one rule:
\[
\compltable{\dl} \triangleq \forall x_1,\ldots,x_m.
\bigvee_{k \in \set{1,\ldots,p}}
\trig{r_k}{x_1,\ldots,x_m}
\]

\smallskip
\noindent
\textbf{Hit policies.}
%As in the case of completeness, when declaring that a table obeys to a
%given single hit policy, there is no guarantee
%that the specified rules are compatible with such a policy.
We start with the unique hit policy, which requires that each
input configuration triggers at most one rule. This can be formalized
as follows:
\[
\uniquetable{\dl} \triangleq \forall \vec{x}.
%\hspace{-.5cm}
\bigwedge_{i \in \set{1,\ldots,p}}
%\hspace{-.5cm}
\trig{r_i}{\vec{x}}
\rightarrow
%\hspace{-.5cm}
\bigwedge_{j \in \set{1,\ldots,p}\setminus\set{i}}
%\hspace{-.5cm}
\neg \trig{r_j}{\vec{x}}
\]

We then continue with the any hit policy. Here multiple rules may be
triggered by the same input configuration, but if so, then they must
agree on the output. This can be formalized as follows:
\[
\anytable{\dl} \triangleq
\hspace{-.5cm}
\bigwedge_{i,j \in \set{1,\ldots,p},i\neq j}
\hspace{-.5cm}
\forall \vec{x} \forall \vec{y}.
\begin{array}[t]{l}
\trig{r_i}{\vec{x}} \\
{}\land \trig{r_j}{\vec{x}}
\rightarrow
\begin{array}[t]{l}
\iorel{r_i}{\vec{x}}{\vec{y}}\\
{} \land \iorel{r_j}{\vec{x}}{\vec{y}}
\end{array}
\end{array}
\]

% Under both the unique and any hit policy, the relationship between
% input and output configurations can be lifted from the single-rule
% case discussed above to the whole decision table as follows:
% \[
% \uaiorel{\dt}{\vec{x}}{\vec{y}} \triangleq
% \bigvee_{r \in R} \iorel{r}{\vec{x}}{\vec{y}}
% \]

We now consider the case of priority hit policy. This requires
to reformulate the rule semantics, so as to consider the
whole decision table and the priority of the rules. In
particular, with this hit policy a rule $r \in \trules$ is
\emph{triggered with priority} by an input
configuration $\vec{x}$ if it is triggered by $\vec{x}$ in the sense specified
above, and no rule of higher priority is triggered by the same input $\vec{x}$:
\[
\ptrig{r}{\vec{x}} \triangleq \trig{r}{\vec{x}} \land
\hspace{-2cm}
\bigwedge_{r_h
  \in \set{r' \mid r' \in \trules \text{ and } \tpr(r') > \tpr(r)}}
\hspace{-2cm}
\neg \trig{r'}{\vec{x}}
\]
With this policy, the relationship between
input and output configurations can be lifted from the single-rule
case discussed above to the whole decision table by isolating the
highest-priority rule that matches with the input configuration, and
by considering its output:
\[
\piorel{\dt}{\vec{x}}{\vec{y}} \triangleq
\bigvee_{r \in R}
\ptrig{r}{\vec{x}} \land
\iorel{r}{\vec{x}}{\vec{y}}
\]
Finally, we observe that the priority hit policy may create a
situation in which some rules are never triggered. This happens when
other rules of higher priority have more general input
conditions. We formalize this notion by introducing a formula
dedicated to check when a rule $r_1 \in \trules$ \emph{is masked} by
another rule $r_2 \in \trules$:
\[
\maskedby{r_1}{r_2} \triangleq \tpr(r_2) > \tpr(r_1) \land \forall
\vec{x}. \trig{r_1}{\vec{x}} \rightarrow \trig{r_2}{\vec{x}}
\]

\vspace{-0.3cm}

\smallskip
\noindent
\textbf{Correctness formula.}
We now combine the previously defined formulae into a single formula
that captures the overall correctness
of a decision table.

We say that $\dt$ is \emph{correct} if the following conditions hold:
\begin{compactenum}
\item Every table cell, i.e., every input condition or
  output object, is legal for the corresponding attribute (considering
  the attribute type and facet).
\item The completeness indicator corresponds to $\cpol$ iff the table is indeed complete.
\item The rules are compatible with the hit policy indicator:
\begin{compactenum}
\item if the hit policy is $\upol$, each input configuration
  triggers at most one rule;
\item if the hit policy is $\apol$, all overlapping rules (i.e.,
  rules that could simultaneously trigger) have the same output;
\item if the hit policy is $\ppol$, all rules are ``useful'',
  i.e., no rule is masked by a rule with higher priority.
\end{compactenum}
\end{compactenum}
Based on the previously introduced formulae, we formalize
correctness as:
\begin{align}
\correct{\dt}
\triangleq
&
\phantom{\land{}}
\bigwedge_{\tup{\incond,\outval} \in
  \trules} \left(
\bigwedge_{\attr{a}\in\tin} \comp{\attr{a}}{\incond(\attr{a})} \land
\bigwedge_{\attr{b}\in\tout} \comp{\attr{b}}{\outval(\attr{b})}
\right)
\tag{1}\\
&\land
\big((\tcompl = \cpol)
\leftrightarrow
\compltable{\dt}\big)
\tag{2}\\
&\land
\big((\thit = \upol)
\rightarrow
\uniquetable{\dt}\big)
\tag{3a}\\
&\land
\big((\thit = \apol)
\rightarrow
\anytable{\dt}\big)
\tag{3b}\\
&\land
\Big((\thit = \ppol)
\rightarrow
\bigwedge_{r_1,r_2 \in \trules}
\neg \maskedby{r_1}{r_2}
\Big)
\tag{3c}
\end{align}

\smallskip
\noindent
\textbf{Global input-output formula.}
We combine the previously defined formulae into a single formula
that captures the overall input-output relation induced by
$\dt$. This is done by exploiting the notion of input-output
related configurations by a rule, so as to cover the entire
table. Specifically we say that an input configuration $\vec{x}$ and
an output configuration $\vec{y}$ are \emph{input-output related} by
$\dt$ if:
\begin{compactenum}
\item the hit policy is either $\upol$ or $\apol$, and there
  exists a rule that relates $\vec{x}$ to $\vec{y}$ (in the case of
  any hit policy, there could be many, but they establish the same
  input-output relation, so it is sufficient to pick one of them);
\item the hit policy is $\ppol$, and there exists a rule relating
  $\vec{x}$ to $\vec{y}$ without any other rule of higher priority
  that is triggered by $\vec{x}$ (is such a rule exists, then it is
  such rule that has to be selected to relate input-output).
\end{compactenum}
This is formalized as follows:
\begin{align}
\iorel{\dt}{\vec{x}}{\vec{y}}
\triangleq
&
\Big(
  (\thit = \upol \lor \thit = \apol)
\rightarrow
\bigvee_{r \in \trules}
\iorel{r}{\vec{x}}{\vec{y}}
\Big)
\tag{1}\\
&
\hspace{-1cm}
\land
\left(
\begin{array}[t]{@{}l@{}l@{}}
(\thit = \ppol)
\rightarrow
\bigvee_{r=\tup{\incond,\outval} \in \trules}
&
\ptrig{r}{\vec{x}}
\\
&\land \bigwedge_{j \in \set{1,\ldots,n}} \matches{\attr{b_j}}{\outval(\attr{b_j})}{y_j}
\end{array}
\right)
  \tag{2}
\end{align}

\vspace{-0.7cm}

% \begin{figure}[t]
%     \centering
%     \includegraphics[width=.6\textwidth]{imgs/Decision_Table_2}
%     \caption{Example decision table}
%     \label{fig:DecisionTable}
% \end{figure}

\section{Algorithms}
\label{sec:algo}

We now introduce algorithms to handle the two main analysis tasks introduced in the previous section: detecting overlapping rules and (in)completeness.
The proposed algorithms rely on a geometric interpretation of a DMN table. Every rule in a table is seen as an iso-oriented hyper-rectangle in an N-dimensional space (where N is a number of columns). Indeed, an input entry in a rule can be seen a constraint over one of the columns (i.e.\ dimensions).
In the case of a numerical column, an input entry is an interval (potentially with an infinite upper or lower bound) and thus it defines a segment or line over the dimension corresponding to that column. In the case of a categorical column, we can map each value of the column's domain to a disjoint interval -- e.g. ``Refinancing'' to [0..1), ``Card payoff'' to [1..2), ``Car leasing'' to [2..3), etc. -- and we can see an input entry under this column as defining a segment (or set of segments) over the dimension corresponding to the column in question.  The conjunction of the entries of a row hence defines a hyper-rectangle, or potentially multiple hyper-rectangles in the case of a multi-valued categorical input entry  (e.g. \{``Refinancing'', ``Car leasing''\}). The hyper-rectangles are iso-oriented because only constraints of the form ``attribute operator literal'' are allowed in S-FEEL and such constraints define iso-oriented lines or segments.

\begin{figure}[hbt]
    \centering
    \includegraphics[width=.5\textwidth]{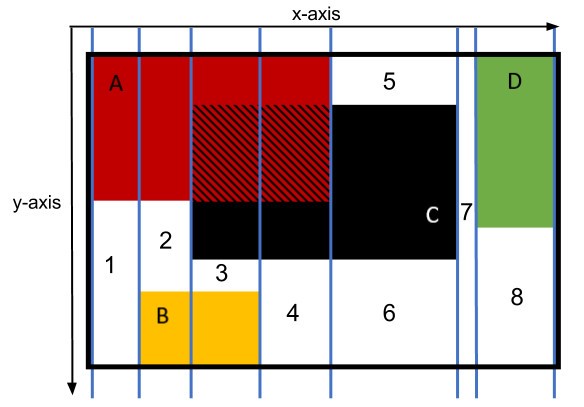}
    \caption{Geometric representation of the DMN table shown in Table 1}
    \label{fig:2D_Example}
   \vspace*{-3mm}
\end{figure}

For example,  the geometric interpretation of Table 1 is shown in \figurename~\ref{fig:2D_Example}. The two dimensions, $x$ and $y$, represent the two input columns (\emph{Annual income} and \emph{Loan size}) respectively. The table contains 4 rules: $A$, $B$, $C$, and $D$. Some of them are overlapping. For example, rule $A$ overlaps with rule $C$. Their intersection is the rectangle $[500,1000]\times[500,1000]$. The table also contains missing values. For example, vector $\langle200, 2000\rangle$ does not match any rule in Table 1.

The algorithms are presented for numeric columns. Minor adaptations (not discussed here) allow these algorithms to handle categorical columns as well. 
%Specifically, in the case of a categorical column, we reason in terms of sets of categorical values and the sorting performed in the algorithms is based on the inclusion relation between sets.

\tikzstyle{leftnode}=[circle,ultra thick,draw=black!75,text width=1mm,inner sep=0pt]
\tikzstyle{rightnode}=[leftnode,fill=black!75]

\subsection{Finding Overlapping Rules}

Algorithm \ref{alg:overlap} finds overlapping rules in a DMN table. This algorithm is an extension of line-sweep algorithm for two-dimensional spatial joins proposed in~\cite{Arge1998}. The idea of this latter algorithm is to pick one dimension (e.g. x-axis), project all objects into this dimension, and then sweep an imaginary line orthogonal to this axis (i.e. parallel to the y-axis). The line stops at every point in the x-axis where either an object starts or ends. When the line makes a ``stop'', we gather all objects that intersect the line (the \emph{active list}). These objects overlap along their x-axis projection. In~\cite{Arge1998}, it is then checked if the objects also overlap in the y-axis, and if so they are added to the result set (i.e.\ the objects overlap). Algorithm \ref{alg:overlap} extends this idea to N dimensions.
The algorithm takes as input:
\begin{compactenum}
  \item ruleList, containing all rules of the input DMN table;
  \item i, containing the index of the column under scrutiny;
  \item N, representing the total number of columns;
  \item OverlappingRuleList, storing the rules that overlap.
\end{compactenum}

The algorithm starts analyzing the first column of the table (axis $x$). All rules are projected over this column. Note that the projection of a rule on a column is an interval.
% \figurename~\ref{fig:Projectionx} shows the projection of the table in on $x$.
We indicate the projection of rule $K$ over axes $x$ and $y$ with $I^{x}_{K}$ and $I^{y}_{K}$ respectively. All the intervals are represented in terms of upper and lower bounds. The bounds are sorted in ascending order (line 7). The algorithm iterates over the list of sorted bounds (line 8).
In the case of \figurename~\ref{fig:2D_Example}, the rules projected over the $x$ axis correspond are:

\begin{center}
\scalebox{0.8}{
\begin{tikzpicture}[node distance=1mm and 8mm]
\node[leftnode] (la) at (0,0) {};
\node[rightnode] (ra) at (3,0) {};
\node[leftnode] (lb) at (.5,-1) {};
\node[rightnode] (rb) at (2.5,-1) {};
\node[leftnode] (lc) at (1.5,-.5) {};
\node[rightnode] (rc) at (4.5,-.5) {};
\node[leftnode] (ld) at (5,.5) {};
\node[rightnode] (rd) at (6.5,.5) {};

\draw[-] (la) edge node[above] {A} (ra);
\draw[-] (lb) edge node[below] {B} (rb);
\draw[-] (lc) edge node[below right] {C} (rc);
\draw[-] (ld) edge node[above] {D} (rd);

\node (lower) at (-1,.5) {lower bound};
\node (upper) at (3.5,.5) {upper bound};
\draw[->] (lower) edge (la);
\draw[->] (upper) edge (ra);
\end{tikzpicture}
}
\end{center}

% \begin{figure}[t]
%     \centering
%     \includegraphics[width=.7\textwidth]{imgs/1Dexample}
%     \caption{Projection of the table in \figurename~\ref{fig:2D_Example} on $x$}
%     \label{fig:Projectionx}
% \end{figure}

Considering the rules above, the algorithm first analyzes the lower bound of $I^{x}_{A}$. Therefore, $I^{x}_{A}$ is added to an active list of intervals for the first column $x$, $\mathcal{L}_{x}$, since the bound processed is a lower bound (line 13). Next, the algorithm processes the lower bound of $I^{x}_{B}$ and $I^{x}_{B}$ is added to $\mathcal{L}_{x}$. Then, the lower bound of $I^{x}_{C}$ is processed and $I^{x}_{C}$ is added to $\mathcal{L}_{x}$. Finally, the algorithm processes the upper bound of $I^{x}_{B}$. Every time an upper bound of an interval is processed (line 9), the following column of the table is analyzed (in this case $y$) by invoking \emph{findOverlappingRules} recursively (line 10).

% \begin{figure}[t]
%     \centering
%     \includegraphics[width=.7\textwidth]{imgs/1Dexample2}
%     \caption{Projection of the table in \figurename~\ref{fig:2D_Example} on $x$}
%     \label{fig:Projectionx2}
% \end{figure}

All the intervals projections on $y$ of the rules corresponding to intervals contained in $\mathcal{L}_{x}$ (in our example $A$, $B$, and $C$) are represented in terms of upper bounds and lower bounds:
\begin{center}
\scalebox{0.7}{
\begin{tikzpicture}[node distance=1mm and 8mm]
\node[leftnode] (la) at (0,0) {};
\node[rightnode] (ra) at (3,0) {};
\node[leftnode] (lb) at (5,-1) {};
\node[rightnode] (rb) at (7,-1) {};
\node[leftnode] (lc) at (1,-.5) {};
\node[rightnode] (rc) at (4,-.5) {};

\draw[-] (la) edge node[above] {A} (ra);
\draw[-] (lb) edge node[below] {B} (rb);
\draw[-] (lc) edge node[below right] {C} (rc);
\end{tikzpicture}
}
\end{center}

 % (\figurename~\ref{fig:Projectionx2}).
 The bounds are sorted in ascending order. The algorithm iterates over the list of sorted bounds.
%If the bound under examination is a lower bound the corresponding rule is added to an active list. For example,
Considering the intervals above, the algorithm first encounters the lower bound of $I^{y}_{A}$. Therefore, $I^{y}_{A}$ is added to the active list of intervals for the second column $y$, $\mathcal{L}_{y}$. Next, the algorithm processes the lower bound of $I^{y}_{C}$ and adds $I^{y}_{C}$ to $\mathcal{L}_{y}$. Then, the upper bound of $I^{y}_{C}$ is processed. Since there is no other column in the table, this means that all the rules corresponding to the intervals in $\mathcal{L}_{y}$ overlap. At the end of each recursion, the interval corresponding to the current bound is removed from the current active list (line 11). In addition, when the last column of the table is processed (line 1), the algorithm checks whether the identified set of overlapping rules is contained in one of the other sets produced in a previous recursion (lines 3). If this is not the case, the new set of overlapping rules is added to the output list overlappingRuleList (line 4). In this way, the procedure outputs maximal sets of overlapping rules having a non-empty intersection stored in overlappingRuleList (line 16).

\begin{algorithm2e}[t]
\scriptsize
\DontPrintSemicolon \SetAlgoVlined
\KwIn{$ruleList$;  $i$; $N$; $overlappingRuleList$.}
\BlankLine
\eIf{$i$ == $N$} { define current overlap $currentOverlapRules$; /* it contains the list of rules that overlap up to the current point */ ; \;
\If{!$overlappingRuleList$.includes($currentOverlapRules$)}{
$overlappingRuleList$.put($currentOverlapRules$); \;
}
}{
define the current list of bounds $\mathcal{L}_{x_{i}}$;\;
%define the variable containing the last boundary $lastBound$;\;
$sortedListAllBounds$ = $ruleList$.sort($i$);\;
\BlankLine
\ForEach{$currentBound \in sortedListAllBoundaries$}{
\eIf{!$currentBound$.isLower()}{
findOverlappingRules($\mathcal{L}_{x_{i}}$,$i$ +1, $N$, $overlappingRuleList$); /* recursive call */\;
$\mathcal{L}_{x_{i}}$.delete($currentBound$);\;
}{
$\mathcal{L}_{x_{i}}$.put($currentBound$);\;
}
$lastBound$ = $currentBound$;\;
}
}
\BlankLine
\Return{$overlappingRuleList$};\;
\caption{Procedure findOverlappingRules. \label{alg:overlap}}
\end{algorithm2e}
\vspace{-0.3cm}

\subsection{Finding Missing Rules}
%The pseudo-code of procedure \emph{findMissingRules} for  in an input DMN table is shown in
Algorithm \ref{alg:missing} describes the procedure for finding missing rules, which is also based on the line-sweep principle. The algorithm takes as inputs 5 parameters:
\begin{compactenum}
  \item ruleList, containing all rules of the input DMN table;
  \item missingIntervals, storing the current missing intervals;
  \item i, containing the index of the column under scrutiny;
  \item N, representing the total number of columns;
  \item MissingRuleList, storing the missing rules.
\end{compactenum}
The algorithm starts analyzing the first column of the table (axis $x$). Consider again the projection of the table in \figurename~\ref{fig:2D_Example} on $x$:

\begin{center}
\scalebox{0.7}{
\begin{tikzpicture}[node distance=1mm and 8mm]
\node[leftnode] (la) at (0,0) {};
\node[rightnode] (ra) at (3,0) {};
\node[leftnode] (lb) at (.5,-1) {};
\node[rightnode] (rb) at (2.5,-1) {};
\node[leftnode] (lc) at (1.5,-.5) {};
\node[rightnode] (rc) at (4.5,-.5) {};
\node[leftnode] (ld) at (5,.5) {};
\node[rightnode] (rd) at (6.5,.5) {};

\draw[-] (la) edge node[above] {A} (ra);
\draw[-] (lb) edge node[below] {B} (rb);
\draw[-] (lc) edge node[below right] {C} (rc);
\draw[-] (ld) edge node[above] {D} (rd);
\end{tikzpicture}
}
\end{center}
Upper and lower bounds of each interval are sorted in ascending order (line 3). The algorithm iterates over the list of sorted bounds (line 4).

Considering the rules above, the algorithm first analyzes the lower bound of $I^{x}_{A}$. Therefore, $I^{x}_{A}$ is added to an active list of intervals for the first column $x$, $\mathcal{L}_{x}$. An interval is added to the active list only if its lower bound is processed (line 15). If the upper bound of an interval is processed, the interval is removed from the list (line 17). Next, the algorithm processes the lower bound of $I^{x}_{B}$. Since $\mathcal{L}_{x}$ is not empty, $I^{x}_{B}$ is not added to $\mathcal{L}_{x}$ yet (line 11). Starting from the interval $\mathcal{I_{A,B}}$ (line 12) having the lower bound of $I^{x}_{A}$ as lower bound and the lower bound of $I^{x}_{B}$ as upper bound, the following column of the table is analyzed (in this case $y$) by invoking \emph{findMissingRules} recursively (line 13).
%
%
%
% \begin{figure}[t]
%     \centering
%     \includegraphics[width=.7\textwidth]{imgs/1Dexample3}
%     \caption{Projection of the table in \figurename~\ref{fig:2D_Example} on $x$}
%     \label{fig:Projectionx3}
% \end{figure}
%
%
All the interval projections on $y$ of the rules corresponding to intervals contained in $\mathcal{L}_{x}$ (in our example only $A$) are represented in terms of upper and lower bounds, obtaining in this case the following simple situation:
\begin{center}
\scalebox{0.7}{
\begin{tikzpicture}[node distance=1mm and 8mm]
\node[leftnode] (la) at (0,0) {};
\node[rightnode] (ra) at (3,0) {};
\draw[-] (la) edge node[above] {A} (ra);
\end{tikzpicture}
}
\end{center}
 % (\figurename~\ref{fig:Projectionx3}).
The bounds are sorted in ascending order. The algorithm iterates over the list of sorted bounds. The first bound taken into consideration is the lower bound of $I^{y}_{A}$ so that $I^{y}_{A}$ is added to $\mathcal{L}_{y}$ (since $\mathcal{L}_{y}$ is empty). Since this bound corresponds to the minimum possible value for $y$, there are no missing values between the minimum possible value for $y$ and the lower bound of $I^{y}_{A}$ (line 5). Next, the algorithm processes the second bound in $\mathcal{L}_{y}$ that is the upper bound of $I^{y}_{A}$. Considering that the upper bound of $I^{y}_{A}$ is the last one in $\mathcal{L}_{y}$, the algorithm checks if this value corresponds to the maximum possible value for $y$ (line 5). Since this is not the case, this means that there are missing values in the area between the upper bound of  $I^{y}_{A}$ and the next bound over the same column (in this case area 1). The algorithm checks if the identified area is contiguous to an area of missing values previously found (line 7). If this is the case the two areas are merged (line 8). If this is not the case, the area is added to a list of missing value areas (line 10). In our case, area 1 is added to a list of missing value areas.
Note that the algorithm merges two areas of missing values only when the intervals corresponding to one column are contiguous and the ones corresponding to all the other columns are exactly the same. In the example in \figurename~\ref{fig:2D_Example}, areas 4 and 6 are merged.

At this point, the recursion ends and the algorithm proceeds analyzing the intervals in the projection along the $x$ axis. % \figurename~\ref{fig:Projectionx}
 The last bound processed was the lower bound of $I^{x}_{B}$, so that $I^{x}_{B}$ is added to $\mathcal{L}_{x}$. Next, the algorithm processes the lower bound of $I^{x}_{C}$ (since $\mathcal{L}_{x}$ is not empty, $I^{x}_{C}$ is not added to $\mathcal{L}_{x}$ yet). Starting from the interval $\mathcal{I_{B,C}}$ having the lower bound of $I^{x}_{B}$ as lower bound and the lower bound of $I^{x}_{C}$ as upper bound, the following column of the table is analyzed (in this case $y$) again through recursion.

% \begin{figure}[t]
%     \centering
%     \includegraphics[width=.7\textwidth]{imgs/1Dexample4}
%     \caption{Projection of the table in \figurename~\ref{fig:2D_Example} on $x$}
%     \label{fig:Projectionx4}
% \end{figure}

All intervals projections on $y$ of the rules corresponding to intervals contained in $\mathcal{L}_{x}$ (in this case $A$ and $B$) are represented in terms of upper and lower bounds:
\begin{center}
\scalebox{0.7}{
\begin{tikzpicture}[node distance=1mm and 8mm]
\node[leftnode] (la) at (0,0) {};
\node[rightnode] (ra) at (3,0) {};
\node[leftnode] (lb) at (4,0) {};
\node[rightnode] (rb) at (6,0) {};
\draw[-] (la) edge node[above] {A} (ra);
\draw[-] (lb) edge node[above] {B} (rb);
\end{tikzpicture}
}
\end{center}
 % (\figurename~\ref{fig:Projectionx4}).
 The bounds are sorted in ascending order. The algorithm iterates over the list of sorted bounds.
%If the bound under examination is a lower bound the corresponding rule is added to an active list. For example,
Considering the rules above, the algorithm first processes the lower bound of $I^{y}_{A}$ so that $I^{y}_{A}$ is added to $\mathcal{L}_{y}$ ($\mathcal{L}_{y}$ is empty). Then, the upper bound of $I^{y}_{A}$ is processed. When the algorithm reaches the upper bound of an interval in a certain column the interval is removed from the corresponding active list. Therefore, $I^{y}_{A}$  is removed from $\mathcal{L}_{y}$. Next, the lower bound of $I^{y}_{B}$ is processed. Since $\mathcal{L}_{y}$ is empty, the algorithm checks if the previous processed bound is contiguous with the current one (line 5). Since this is not the case, this means that there are missing values in the area between the upper bound of $I^{y}_{A}$ and the next bound over the same column (in this case area 2). The algorithm checks if the identified area is contiguous to an area of missing values previously found. If this is the case, the two areas are merged. If this is not the case, the area is added to a list of missing value areas (in our case area 2 is added to a list of missing value areas). The list of missing areas stored in missingRuleList is returned by the algorithm (line 19).

\begin{algorithm2e}[t]
\scriptsize
\DontPrintSemicolon \SetAlgoVlined
\KwIn{$ruleList$; $missingIntervals$; $i$; $N$; $missingRuleList$.}
\BlankLine
\If{$i$ $>$ $N$} {
define the current list of boundaries $\mathcal{L}_{x_{i}}$;\;
%define the variable containing the last boundary $lastBound$;\;
$sortedListAllBoundaries$ = $ruleList$.sort($i$);\;
%define the list containing the possible values for a categorical type $allowedCategoricalValues$;\;
%\If{$ruleList$.isColumnTypeCategorical($i$}{
%$allowedCategoricalValues$ = getAllowedCategoricalValuesFromColumn($i$);\;
%}
\ForEach{$currentBound \in sortedListAllBoundaries$} {

%/* it is equal to true if the there is a missing interval between the previous boundary and the current boundary */  \;
%define the variable $missingInterval$ containing the missing interval between the previous boundary and the current boundary  \;

%\uIf{$currentBound$.isFirst()~ \&\&~ $currentBound$.getValue() != -Infinity}{
%$missingInterval$ = constructInterval(-Infinity, $currentBound$);\;
%$isMissingInterval$ = true;\;
%}\uElseIf{$currentBound$.isLast()~ \&\& ~$currentBound$.getValue() != Infinity}
%{
%$missingInterval$ = constructInterval($currentBound$, Infinity);\;
%$isMissingInterval$ = true;\;

%}\uElse{

\If{!areContiguous($lastBound$, $currentBound$)}{
$missingIntervals$[$i$] = constructInterval($lastBound$, $currentBound$);\;
\eIf{$missingRuleList$.canBeMerged($missingIntervals$);}{
$missingRuleList$.merge($missingIntervals$);\;
%}
}{
$missingRuleList$.add($missingIntervals$);\;
}
}
%/* it is equal to true if the current boundary is part of a new interval with respect to the previous boundary */  \;
\If{!$\mathcal{L}_{x_{i}}$.isEmpty() )}{
$missingIntervals$ [$i$] = constructInterval($lastBound$, $currentBound$);\;

findMissingRules($\mathcal{L}_{x_{i}}$,$missingIntervals$,$i$ +1, $N$, $missingRuleList$); /* recursive invocation */\;

}
%$currentKey$ = $currentBound$.getKey();\;
\eIf{$currentBound$.isLower()}{
$\mathcal{L}_{x_{i}}$.put($currentBound$);\;
}{
$\mathcal{L}_{x_{i}}$.delete($currentBound$);\;
}
$lastBound$ = $currentBound$;\;
}
}
\BlankLine
\Return{$missingRuleList$};\;
\caption{Procedure findMissingRules. \label{alg:missing}}
\end{algorithm2e}

\vspace{-0.3cm}

\section{Evaluation}
\label{sec:exper}

We implemented the algorithms on top of \textsf{dmn-js}: the open-source rendering and editing toolkit of Camunda DMN.\footnote{\url{https://camunda.org/}} In it current version, \textsf{dmn-js} does not support correctness verification. Our \textsf{dmn-js} extension with verification features can be found at \url{https://github.com/ulaurson/dmn-js} and a deployed version is available for testing at \url{http://kodu.ut.ee/~ulaurson/DMN/}.

%Camunda is an open source platform that can be used for workflow and business process management.

For the evaluation, we created decision tables from a loan dataset of LendingClub -- a peer-to-peer lending marketplace.\footnote{Dataset available at \url{https://www.lendingclub.com/info/download-data.action}} The employed dataset contains data about all loans issued in 2013-2014 (23~5629 loans). For each loan, there are attributes of the loan itself (e.g., amount, purpose), of the lender (e.g., income, family status, property ownership), and a credit grade (A, B, C, D, E, F, G).

%including the current loan status (Current, Late, Fully Paid, etc.) and latest payment information. In particular, it contains

%In this peer-to-peer marInvestors provide the capital to enable loans in exchange for earning interest. Borrowers access lower interest rate loans through an online or mobile interface.

Using Weka~\cite{Weka}, we trained decision trees to classify the  grade of each loan from a subset of the loan attributes. We then translated each trained decision tree into a DMN table by mapping each path from the root to a leaf of the tree into a rule. Using different attributes and pruning parameters in the decision tree discovery, we generated DMN tables containing approx.\ 500, 1000 and 1500 rules and 3, 5 and 7 columns (nine tables in total). The 3-dimensional (i.e. 3-column) tables have one categorical and two numerical input columns; the 5-dimensional tables have two categorical and three numerical input columns, and the 7-dimensional tables has two categorical and five numerical input columns.

%The DMN tables represent business rules for determining the loan grade. In total, there are 7 different values for loan grade:  A, B, C, D, E, F,and G. \todo{Can we change these values?}
%Therefore, we generated 9 DMN tables in total.

By construction, the generated tables do not contain overlapping or missing rules.
To introduce missing rules in a table, we selected $10\%$ of the rules. For each of them, we then randomly selected one column, and we injected noise into the input entry in the cell in the selected column by decreasing its lower bound and increasing its upper bound in the case of a numerical domain (e.g.\ interval [3..6] becomes [2..7]) and by adding one value in the case of a categorical domain (e.g. \{ Refinancing, CreditCardPayoff \} becomes \{ Refinancing, CreditCardPayoff, Leasing \}). These modifications make it that the rule will overlap others. Conversely, to introduce missing rule errors, we selected $10\%$ of the rules, picked a random column for each row and ``shrank'' the corresponding input entry.

%Research questions:

%is the algorithm to derive the set of overlapping rules in a DMN table effective?
%is the algorithm to derive the set of missing rules in a DMN table effective?
%is the algorithm to derive the set of overlapping rules in a DMN table efficient?
%is the algorithm to derive the set of missing rules in a DMN table efficient?

We checked each generated table both for missing and incomplete rules and measured execution times averaged over 5 runs on a single core of a 64-bit 2.2 Ghz Intel Core i5-5200U processor with 16GB of RAM. The results are shown in \tablename~\ref{tab:exp-time}. Execution times for missing rules detection are under 2 seconds, except for the 7-columns tables with 1000-1500 rules. The detection of overlapping rules leads to higher execution times, due to the need to detect sets of overlapping rules and ensure maximality. The execution times for overlapping rules detection on the 3-columns tables is higher than on the 5-columns tables because the 5-columns tables have less rule overlaps. This is because there are proportionally less categorical columns in the 5-columns tables than in the 3-columns ones, and the modifications made to categorical columns create more overlaps.

%The implementation is written in JavaScript and runs on the NodeJS runtime environment.

In addition to implementing our algorithms, we implemented algorithms designed to produce the same output as Signavio. In Signavio, if multiple rules have a joint intersection (e.g. rules \{r1, r2, r3\}) the output contains an overlap entry for the triplet \{r1, r2, r3\} but also for the pairs \{r1, r2\}, \{r2, r3\} and \{r1, r3\} (i.e. subsets of the overlapping set). Furthermore, the overlap of pair \{r1, r2\} may be reported multiple times if r3 breaks $r1 \cap r2$ into multiple hyper-rectangles (and same for \{r2, r3\} and \{r1, r3\}).
Meanwhile, our approach produces only maximal sets of overlapping rules with a non-empty intersection.
%In contrast, \theirs~reports the same rule overlap multiple times in cases where more than two rules have a joint non-empty intersection.
%To evaluate the effectiveness of \ours, we compare it with the approach implemented in \theirs. In particular,

\tablename~\ref{tab:exp-comp} shows the number of sets of overlapping rules and the number of missing rules identified by \ours~vs.\ Signavio's one. In all runs, both the number of overlapping and missing rules is drastically lower in \ours.

\begin{table}[hbt]
\scriptsize
\begin{tabularx}{\textwidth}{|l|R|R|R|R|R|R|R|R|R|}
\cline{2-10}
\multicolumn{1}{c|}{}
&
\multicolumn{3}{c|}{\textsc{3 columns}}
&
\multicolumn{3}{c|}{\textsc{5 columns}}
&
\multicolumn{3}{c|}{\textsc{7 columns}}
\\
\hline
\#rules&
499&998&1\,492&
505&1\,000&1\,506&
502&1\,019&1\,496\\
\hline
overlapping time&
297ms&6\,475ms&24\,530ms&
200ms&1\,621ms&5\,374ms&
5\,715ms&6\,793ms&30\,736ms\\
\hline
missing time&
160ms&611ms&1\,672ms&
163ms&820ms&1\,942ms&
2\,173ms&7\,029ms&18\,263ms\\
\hline
\end{tabularx}
\caption{Execution times (in milliseconds)}
\label{tab:exp-time}
\end{table}

\begin{table}[hbt]
\scriptsize
\begin{tabularx}{\textwidth}{|p{2cm}|l|R|R|R|R|R|R|R|R|R|}
\cline{3-11}
\multicolumn{2}{c|}{}
&
\multicolumn{3}{c|}{\textsc{3 columns}}
&
\multicolumn{3}{c|}{\textsc{5 columns}}
&
\multicolumn{3}{c|}{\textsc{7 columns}}
\\
\hline
\multicolumn{2}{|l|}{\#rules}&
499&998&1\,492&
505&1\,000&1\,506&
502&1\,019&1\,496\\
\hline
\#overlapping&\ours&
131&447&812&
110&225&378&
139&227&371\\
\cline{2-11}
\hspace{.3cm}rule sets&\theirs&
1\,226&10\,920&23\,115&
679&3\,692&8\,921&
23\,175&22\,002&62\,217\\
\hline
\#missing&\ours&
117&330&726&
136&254&462&
134&322&518\\
\cline{2-11}
\hspace{.3cm}rules&\theirs&
668&2\,655&5\,386&
563&2\,022&4\,832&
5\,201&18\,076&43\,552\\
\hline
\end{tabularx}
\caption{Number of reported errors of type ``overlapping rules'' \& ``missing rule''}
\label{tab:exp-comp}
\end{table}

\vspace{-0.6cm}

\section{Conclusion and Future Work}
\label{sec:conclusion}

This paper presented a formal semantics of DMN decision tables, a notion of DMN table correctness, and algorithms that operationalize two core elements of this correctness notion: the detection of overlapping rules and  of missing rules. The algorithms have been implemented atop the DMN toolkit \textsf{dmn-js}. An empirical evaluation on large decision tables has shown the potential for scalability of the proposed algorithms and their ability to generate non-redundant feedback that is more concise than the one generated by the Signavio DMN editor.

The proposed algorithms rely on a geometric interpretation of rules in decision tables, which we foresee could be used to tackle other analysis problems. In particular, we foresee that the problem of simplification of decision tables (rule merging) could be approached from a geometric standpoint. Indeed, if we see the rules as hyperrectangles, the problem of table simplification can be mapped to one of finding an optimal way of merging hyperrectangles with respect to some optimality notion. Another direction for future work is to extend the proposed formal semantics to encompass other aspects of the DMN standard, such as the concept of Decision Requirements Graphs (DRGs), which allow multiple decision tables to be linked in various ways.
%Finally, while the current proposal focuses on analyzing DMN tables where the expressions are written in S-FEEL (i.e., expressions of the form attribute-operator-literal), future work shall focus on supporting more complex types of expressions.

\smallskip\noindent\textbf{Acknowledgement.} This research was partly funded by an Institutional Grant of the Estonian Research Council.


\begin{thebibliography}{10}

\bibitem{Arge1998}
Lars Arge, Octavian Procopiuc, Sridhar Ramaswamy, Torsten Suel, and J{effrey
  Scott} Vitter.
\newblock Scalable sweeping-based spatial join.
\newblock In {\em VLDB}, 1998.

\bibitem{Batoulis2015}
Kimon Batoulis, Andreas Meyer, Ekaterina Bazhenova, Gero Decker, and Mathias
  Weske.
\newblock Extracting decision logic from process models.
\newblock In {\em CAiSE'15}.

\bibitem{CodasylReport}
{CODASYL Decision Table Task Group}.
\newblock {\em A Modern appraisal of decision tables : a {CODASYL} report}.
\newblock ACM, 1982.

\bibitem{Z3}
Leonardo~Mendon{\c{c}}a de~Moura and Nikolaj Bj{\o}rner.
\newblock {Z3:} an efficient {SMT} solver.
\newblock In {\em Proc. of {TACAS}}, pages 337--340. Springer, 2008.

\bibitem{Weka}
Mark~A. Hall, Eibe Frank, Geoffrey Holmes, Bernhard Pfahringer, Peter
  Reutemann, and Ian~H. Witten.
\newblock The {WEKA} data mining software: an update.
\newblock {\em {SIGKDD} Explorations}, 11(1):10--18, 2009.

\bibitem{DMN}
{Object Management Group}.
\newblock {Decision Model and Notation (DMN) 1.0}, 2015.

\bibitem{Pawlak1987}
Zdzislaw Pawlak.
\newblock Decision tables -- a rough set approach.
\newblock {\em Bulletin of the EATCS}, 33:85--95, 1987.

\bibitem{Pooch74}
Udo~W. Pooch.
\newblock Translation of decision tables.
\newblock {\em Comp. Surv.}, 6(2):125--151, 1974.

\bibitem{Vanthienen1994}
Jan Vanthienen and Elke Dries.
\newblock Illustration of a decision table tool for specifying and implementing
  knowledge based systems.
\newblock {\em International Journal on Artificial Intelligence Tools},
  3(2):267--288, 1994.

\bibitem{Vanthienen1998}
Jan Vanthienen, Christophe Mues, and Ann Aerts.
\newblock An illustration of verification and validation in the modelling phase
  of {KBS} development.
\newblock {\em Data Knowl. Eng.}, 27(3):337--352, 1998.

\end{thebibliography}
\end{document}